\documentclass[11pt,a4paper]{article}

\usepackage{jheppub}
\usepackage{calc}
\usepackage{rotating}
\usepackage[english]{babel}
\usepackage{graphicx}
\usepackage{epsfig,subfig}
\usepackage{float}
\usepackage{amsmath}
\usepackage{amssymb}
\usepackage{amsthm}
\usepackage{latexsym}
\usepackage{dcolumn}
\usepackage{multirow}
\usepackage{color}
\usepackage{hyperref}
\usepackage{mciteplus}
\usepackage{soul}
\usepackage{slashed}

\newcommand{\newc}{\newcommand*}

\long\def\begincomment#1\endcomment{%
        \begingroup\sf\baselineskip12pt#1\endgroup}

\newc{\etal}{\textrm{et al.}} 
\newc{\eg}{\textrm{e.g.}} 
\newc{\ie}{\textrm{i.e.}}
\newc{\etc}{\textrm{etc.}}
\newc\vs{\textrm{vs.}}
\newc{\cl}{\rm {C.L.}}

\newc{\ev}{\ensuremath{\,\mathrm{eV}}}
\newc{\kev}{\ensuremath{\,\mathrm{keV}}}
\newc{\mev}{\ensuremath{\,\mathrm{MeV}}}
\newc{\gev}{\ensuremath{\,\mathrm{GeV}}}
\newc{\tev}{\ensuremath{\,\mathrm{TeV}}}

\newc{\invpb}{\ensuremath{/\text{pb}}}
\newc{\invfb}{\ensuremath{/\text{fb}}}

\newc{\chisqmin}{\ensuremath{\chi^2_{\mathrm{min}}}}
\newc{\delchisq}{\ensuremath{\Delta\chi^2}}
\newc{\chisq}{\ensuremath{\chi^2}}
\newc{\like}{\ensuremath{\mathcal{L}}}

\newc\lsim{\ensuremath{\mathrel{\rlap{\lower4pt\hbox{\hskip1pt$\sim$}}\raise1pt\hbox{$<$}}}}
\newc\gsim{\ensuremath{\mathrel{\rlap{\lower4pt\hbox{\hskip1pt$\sim$}}\raise1pt\hbox{$>$}}}}
\newc{\VEV}[1]{\ensuremath{\langle #1 \rangle}}
\newc{\dl}{\ensuremath{\stackrel{\leftarrow}{D}}}
\newc{\dr}{\ensuremath{\stackrel{\rightarrow}{D}}}

\newc{\bi}{\begin{itemize}}
\newc{\ei}{\end{itemize}}
\newc{\bed}{\begin{description}}
\newc{\eed}{\end{description}}
\newc{\ben}{\begin{enumerate}}
\newc{\een}{\end{enumerate}}

\newc{\be}{\begin{equation}}
\newc{\ee}{\end{equation}}
\newc{\bea}{\begin{eqnarray}}
\newc{\eea}{\end{eqnarray}}

\newc{\alphas}{\ensuremath{\alpha_s}}
\newc{\alphatwo}{\ensuremath{\alpha_2}}
\newc{\alphaone}{\ensuremath{\alpha_1}}
\newc{\alphai}[1]{\ensuremath{\alpha_{#1}}}
\newc{\alphaem}{\ensuremath{\alpha_{\mathrm{em}}}}

\newc{\alphaeff}{\ensuremath{\alpha_{\mathrm{eff}}}}

\newc{\sineff}{\ensuremath{\sin \theta_{\mathrm{eff}}}}
\newc{\sinsqeff}{\ensuremath{\sin^2 \theta_{\mathrm{eff}}}}
\newc{\dalphahad}{\ensuremath{\Delta \alpha_{\mathrm{had}}}}

\newc{\yt}{\ensuremath{h_t}} 
\newc{\yb}{\ensuremath{h_b}} 
\newc{\ytau}{\ensuremath{h_{\tau}}}

\newc\mz{\ensuremath{m_Z}} 
\newc\mw{\ensuremath{m_W}}

\newc{\mtop}{\ensuremath{ m_t}}               
\newc{\mtpole}{\ensuremath{ M_t}}
\newc{\mbottom}{\ensuremath{ m_b}} 
\newc{\mtau}{\ensuremath{ m_{\tau}}}

\newc{\sthw}{\ensuremath{ \sin\theta_W}}              
\newc{\cthw}{\ensuremath{\cos\theta_W}}
\newc{\tanthw}{\ensuremath{ \tan\theta_W}}              
\newc{\cotthw}{\ensuremath{\cot\theta_W}}
\newc{\ssqthw}{\ensuremath{\sin^2 \theta_W}}

\newc\hsm{\ensuremath{H_{\mathrm{SM}}}}
\newc\mhsm{\ensuremath{m_{H_{\mathrm{SM}}}}}
\newc\hobs{\ensuremath{H_{\mathrm{obs}}}}
\newc\mhobs{\ensuremath{m_{H_{\mathrm{obs}}}}}

\newc{\hig}[1]{\ensuremath{H_{#1}}}
\newc{\psu}[1]{\ensuremath{A_{#1}}} 
\newc{\ma}[1]{\ensuremath{m_{A_{#1}}}} 
\newc{\mh}[1]{\ensuremath{m_{H_{#1}}}}

\newc{\hu}{\ensuremath{H_u}} 
\newc{\hd}{\ensuremath{H_d}}
\newc{\hs}{\ensuremath{H_s}}      
\newc{\mhu}{\ensuremath{m_{H_u}}}  
\newc{\mhd}{\ensuremath{m_{H_d}}}
\newc{\mhs}{\ensuremath{m_{H_s}}} 

 \newc{\mhuew}{\ensuremath{m^{\ast}_{H_u}}}       
\newc{\mhdew}{\ensuremath{m^{\ast}_{H_d}}}
 \newc{\mhuewsq}{\ensuremath{m^{\ast\, 2}_{H_u}}}       
\newc{\mhdewsq}{\ensuremath{m^{\ast\, 2}_{H_d}}}

\newc{\robsgg}{\ensuremath{R_{h_{\rm obs}}(\gamma\gamma)}}
\newc{\rtwozz}{\ensuremath{R_{h_{\rm obs}}(ZZ)}}
\newc{\rgg}[1]{\ensuremath{R_{h_{#1}}(\gamma\gamma)}}
\newc{\rzz}[1]{\ensuremath{R_{h_{#1}}(ZZ)}}

\newc{\ptcut}{\ensuremath{\slashed p_T}}
\newc{\met}{\ensuremath{\slashed E_T}}
\newc{\muco}{\ensuremath{\mu_{\rm col}}}

\newc{\lepplepm}{\ensuremath{\ell^+\ell^-}}
\newc{\tauptaum}{\ensuremath{ \tau^+\tau^-}}

\newc{\qqbar}{\ensuremath{ q\bar{q}}} 
\newc{\ppbar}{\ensuremath{ p\bar{p}}}
\newc{\bbbar}{\ensuremath{ b\bar{b}}} 
\newc{\ttbar}{\ensuremath{ t\bar{t}}}
\newc{\ffbar}{\ensuremath{ f\bar{f}}} 
\newc{\tautaubar}{\ensuremath{ \tau\bar{\tau}}}

\newc{\mgut}{\ensuremath{M_{\rm GUT}}}
\newc{\mpl}{\ensuremath{M_{\rm Pl}}}
\newc{\msusy}{\ensuremath{M_{\rm SUSY}}}      
\newc{\astau}{\ensuremath{{A_{\tau}}}}
\newc{\astop}{\ensuremath{{A_t}}}
\newc{\asbot}{\ensuremath{{A_b}}}

\newc{\mzero}{\ensuremath{{M_0}}}
\newc{\mhalf}{\ensuremath{ M_{1/2}}}
\newc{\tanb}{\ensuremath{\tan\beta}}
\newc{\azero}{\ensuremath{A_0}}
\newc{\signmu}{\ensuremath{\rm{sgn}\,\mu}}

\newc{\mueff}{\ensuremath{\mu_{\rm{eff}}}}
\newc{\lam}{\ensuremath{{\lambda}}}
\newc{\kap}{\ensuremath{{\kappa}}}
\newc{\alam}{\ensuremath{{A_{\lambda}}}}
\newc{\akap}{\ensuremath{{A_{\kappa}}}}

\newc{\neut}[1]{\ensuremath{\widetilde{\chi}^0_{#1}}}
\newc{\chgo}[1]{\ensuremath{\widetilde{\chi}^{\pm}_{#1}}}

\newc{\mneut}[1]{\ensuremath{m_{\widetilde{\chi}^0_{#1}}}}
\newc{\mchgo}[1]{\ensuremath{m_{\widetilde{\chi}^{\pm}_{#1}}}}

\newc{\squark}{\ensuremath{\tilde{q}}}
\newc{\slepton}{\ensuremath{\tilde{l}}}
\newc{\gluino}{\ensuremath{\tilde{g}}} 
\newc{\mglu}{\ensuremath{m_{\tilde g}}} 
\newc{\mul}{\ensuremath{m_{\tilde{u}_L}}} 
\newc{\mtone}{\ensuremath{m_{\tilde{t}_1}}} 
\newc{\stau}{\ensuremath{\tilde{\tau}}}

\newc{\msbar}{\ensuremath{\overline{MS}}} \newc{\drbar}{\ensuremath{\overline{DR}}}
\newc{\mtmtsmmsbar}{\ensuremath{ m_t(m_t)^{\msbar}_{{\mathrm{SM}}}}}
\newc{\mtmtsmdrbar}{\ensuremath{ m_t(m_t)^{\drbar}_{{\mathrm{SM}}}}}
\newc{\mtmtmssmdrbar}{\ensuremath{ m_t(m_t)^{\drbar}_{{\mathrm{SUSY}}}}}
\newc{\mbmbmsbar}{\ensuremath{ m_b(m_b)^{\msbar} }}
\newc{\mbmbsmmsbar}{\ensuremath{ m_b(m_b)^{\msbar}_{{\mathrm{SM}}}}}
\newc{\mbmzsmmsbar}{\ensuremath{ m_b(\mz)^{\msbar}_{{\mathrm{SM}}}}}
\newc{\mbmzsmdrbar}{\ensuremath{ m_b(\mz)^{\drbar}_{{\mathrm{SM}}}}}
\newc{\mbmzmssmdrbar}{\ensuremath{ m_b(\mz)^{\drbar}_{{\mathrm{SUSY}}}}}
\newc{\mtaumzsmmsbar}{\ensuremath{ m_{\tau}(\mz)^{\msbar}_{{\mathrm{SM}}}}}
\newc{\mtaumzsmdrbar}{\ensuremath{ m_{\tau}(\mz)^{\drbar}_{{\mathrm{SM}}}}}
\newc{\mtaumzmssmdrbar}{\ensuremath{ m_{\tau}(\mz)^{\drbar}_{{\mathrm{SUSY}}}}}
\newc{\alphasmzms}{\ensuremath{\alpha_s(m_Z)^{\overline{MS}}}}
\newc{\alphaimzms}[1]{\ensuremath{\alpha_{#1}(m_Z)^{\overline{MS}}}}
\newc{\alphaemmz}{\ensuremath{\alpha_{\mathrm{em}}(m_Z)^{\overline{MS}}}}

\newc{\sigsip}{\ensuremath{\sigma^{\rm SI}_{p}}}	
\newc{\sigsin}{\ensuremath{\sigma^{\rm SI}_{n}}}
\newc{\sigsdp}{\ensuremath{\sigma^{\rm SD}_{p}}}	
\newc{\sigsdn}{\ensuremath{\sigma^{\rm SD}_{n}}}
\newc{\sigsi}{\ensuremath{\sigma^{\rm SI}}}	
\newc{\sigsd}{\ensuremath{\sigma^{\rm SD}}}

\newc{\abund}{\ensuremath{ \Omega h^2}}
\newc{\omegadm}{\ensuremath{ \Omega_{{\rm DM}}}}     
\newc{\abunddm}{\ensuremath{ \Omega_{{\rm DM}} h^2}} 
\newc{\omegam}{\ensuremath{ \Omega_{{\rm m}}}}       
\newc{\abundm}{\ensuremath{ \Omega_{{\rm m}} h^2}}
\newc{\omegab}{\ensuremath{ \Omega_{{\rm b}}}}	
\newc{\abundb}{\ensuremath{ \Omega_{{\rm b}} h^2}}
\newc{\omegatot}{\ensuremath{ \Omega_{{\rm TOT}}}}
\newc{\omegacdm}{\ensuremath{ \Omega_{{\rm CDM}}}}
\newc{\abundcdm}{\ensuremath{ \Omega_{{\rm CDM}} h^2}}
\newc{\omegalambda}{\ensuremath{ \Omega_{\Lambda}}}
\newc{\abundlambda}{\ensuremath{ \Omega_{\Lambda} h^2}}
\newc{\omegarad}{\ensuremath{ \Omega_{{\rm rad}}}}  
\newc{\abundrad}{\ensuremath{ \Omega_{{\rm rad}} h^2}}

\newc{\rhocrit}{\ensuremath{ \rho_{\rm crit}}}
\newc{\rhochi}{\ensuremath{ \rho_{\chi}}}

\newc{\abundchi}{\ensuremath{\Omega_{\widetilde{\chi}^0_1} h^2}}
\newc{\abundlsp}{\ensuremath{\Omega_{\rm LSP}h^2}}

\newc{\amu}{\ensuremath{ a_{\mu}}}        
\newc{\amususy}{\ensuremath{ a_{\mu}^{\mathrm{SUSY}}}}
\newc{\amuexpt}{\ensuremath{ a_{\mu}^{\mathrm{expt}}}}        
\newc{\amusm}{\ensuremath{ a_{\mu}^{\mathrm{SM}}}}
\newc{\deltaamu}{\ensuremath{\Delta a_{\mu}}} 
\newc{\deltaamususy}{\ensuremath{\delta a_{\mu}^{\mathrm{SUSY}}}}
\newc{\gmtwo}{\ensuremath{ (g-2)_{\mu}}} 
\newc{\deltagmtwomususy}{\ensuremath{\delta\left(g-2\right)_{\mu}^{\mathrm{SUSY}}}}
\newc{\deltagmtwomu}{\ensuremath{\delta\left(g-2\right)_{\mu}}}

\newc\BR{\ensuremath{\rm BR}}

\newc\bsgamma{\ensuremath{ b\to s \gamma }}
\newc\bxsgamma{\ensuremath{\overline{B}\to X_{s}\gamma}}

\newc\brbsgamma{\ensuremath{\BR\left(\bsgamma\right)}}
\newc\brbxsgamma{\ensuremath{\BR\left(\bxsgamma\right)}}

\newc\bsmumu{\ensuremath{B_s\to\mu^+\mu^-}}
\newc\brbsmumu{\ensuremath{\BR\left(B_s\to\mu^+\mu^-\right)}}

\newc\bdmmumu{\ensuremath{\overline{B}_d\to\mu^+\mu^-}}

\newc\bbbarmix{\ensuremath{\overline{B}_s\mbox{-}B_s}}      
\newc\delmbs{\ensuremath{\Delta M_{B_s}}}

\newc{\butaunu}{\ensuremath{B_u \to \tau \nu}}
\newc{\brbutaunu}{\ensuremath{\BR\left(B_u \to \tau \nu\right)}}

\newc{\ssusy}{{\sc SOFTSUSY}}
\newc{\feynhiggs}{{\sc FeynHiggs}}
\newc{\MTN}{{\sc MultiNest}}
\newc{\herwig}{{\sc Herwig++}}
\newc{\susyhit}{{\sc SUSY-HIT}}
\newc{\pythia}{{\sc Pythia}}
\newc{\xenon}{{\sc XENON100}}
\newc{\suprelic}{{\sc SuperIso Relic}}
\newc{\stauc}{\ensuremath{\stau-{\rm coannihilation}}}

\newc{\THDMC}{{\sc 2hdmc}}
\newc{\FH}{{\sc FeynHiggs}}
\newc{\NMT}{{\sc NMSSMTools}}
\newc{\MO}{{\sc MicrOMEGAs}}
\newc{\MMA}{{\sc Mathematica}}
\newc{\FA}{{\sc FeynArts}}
\newc{\FRU}{{\sc FeynRules}}
\newc{\FC}{{\sc FormCalc}}
\newc{\LOTO}{{\sc LoopTools}}
\newc{\HiB}{{\sc HiggsBounds}}
\newc{\MadG}{{\sc MadGraph}}
\newc{\SI}{{\sc SuperIso}}
\newc{\DELP}{{\sc DELPHES}}
\newc{\Fastj}{{\sc FastJet}}
\newc{\Prosp}{{\sc Prospino}}
\newc{\cmate}{{\sc CheckMATE}}

\subheader{
\textrm{\begin{flushright}
APCTP-PRE2015-010 \\
IPMU15-0052
\end{flushright}}}

\title{$\mathcal{O}$(1)~GeV dark matter in SUSY and a very light pseudoscalar at the LHC}

\author[a]{Chengcheng Han,}
\author[a]{Doyoun Kim,}
\author[a]{Shoaib Munir,}
\author[a,b,c]{and Myeonghun Park}
\affiliation[a]{Asia Pacific Center for Theoretical Physics,\\
San 31, Hyoja-dong, Nam-gu, Pohang 790-784, Republic of Korea}
\affiliation[b]{Department of Physics,\\ Postech, Pohang 790-784, Korea}
\affiliation[c]{Kavli IPMU (WPI),\\ The University of Tokyo, Kashiwa, Chiba 277-8583, Japan}
\emailAdd{hancheng@apctp.org}
\emailAdd{doyoun.kim@apctp.org}
\emailAdd{s.munir@apctp.org}
\emailAdd{parc@apctp.org}

\abstract{We analyze the prospects of the detection of an $\mathcal{O}$(1)\gev\
neutralino dark matter, $\neut{1}$, in the Next-to-Minimal
Supersymmetric Standard
Model at the 14\,TeV LHC. We perform dedicated scans of the 
relevant parameter space of the model and find a large number of
points where the thermal relic abundance due to such a dark matter 
is consistent with the PLANCK measurement. We note that this dark matter is highly 
singlino-dominated and is always accompanied by a pseudoscalar, $A_1$, with a 
mass around twice its own, which is responsible for its resonant
annihilation. For two benchmark points from our scan, we then
carry out a detector-level signal-to-background analysis of the
pair production of a heavier higgsino neutralino and a chargino. The higgsino
thus produced decays into the dark matter and either the $Z$ boson or the
$A_1$. For the $Z$-associated production of \neut{1}, we investigate the 
scope of the trilepton search channel. For the $A_1$-associated
production mode, in order to identify the two collimated muons 
coming from the decay of the $A_1$, we employ an unconventional 
method, of clustering them together into one jet-like object. 
Using this method, for certain parameter space
configurations, a much larger sensitivity can be obtained at the 
14\,TeV LHC for the $A_1\neut{1}$ channel compared to the $Z\neut{1}$
channel, with an integrated luminosity of 300\,fb$^{-1}$.}

\arxivnumber{1504.05085}

\begin{document}
\maketitle

\section{\label{intro}Introduction}
The Minimal Supersymmetric Standard Model (MSSM) contains three
neutral Higgs states. The lightest one of these, $h$, is one of the main
candidates for the Higgs boson, \hobs, observed at the
Large Hadron Collider (LHC)~\cite{Chatrchyan:2012ufa,Aad:2012tfa}. The
MSSM also contains a total of four neutralino mass eigenstates, the
lightest of which, \neut{1}, is an important dark matter (DM)
candidate, when it is also the lightest supersymmetric particle (LSP) and 
$R$-parity is conserved. Despite its being
a theoretically sound and highly predictive model, the absence of any
evidence of supersymmetry (SUSY) during Run-I of the LHC has brought
the MSSM under increased scrutiny recently. This is because
 the only way for $h$ to gain a mass close to the one
measured for the \hobs\ is through large radiative
corrections from the SUSY sector. The most dominant of these
corrections come from the stops, which warrants either TeV-scale
SUSY-breaking masses for them or a
 multi-TeV stop mixing parameter, $A_t$. This in turn leads to excessive
 fine-tuning of the model parameters such as the
 soft masses of the two Higgs doublet fields, $H_u$ and
 $H_d$, and the Higgs-higgsino mass paramater $\mu$, given the
 measured mass of the $Z$ boson.

The parameters of the MSSM get further constrained if one requires, in addition to
$h$ fulfilling the role of the observed Higgs boson, \neut{1} to
generate the observed {\it thermal} DM relic density of the universe,
\omegadm, as measured
by the WMAP~\cite{Hinshaw:2012aka} and PLANCK~\cite{Planck:2015xua}
telescopes. This
  condition is strongly dependent on the interplay between the mass
  and the composition of the \neut{1}, as these quantities govern its annihilation
  rate into pairs of SM particles. The \neut{1} is a pure higgsino when
  $M_1,\,M_2 \gg \mu$, with $M_1$ and $M_2$ being the soft gaugino
  masses, and can only yield the correct relic abundance if $\mu \sim
  1$\tev. If \neut{1}\ is instead a bino-higgsino mixture, it can
  give sufficient \omegadm\ for comparatively smaller
  masses, by annihilating via chargino/neutralino
exchanges into pairs of vector bosons. A predominantly bino-like \neut{1}, on the
  other hand, can be consistent with the measured \omegadm\ for much
smaller masses, provided one of the third-generation sfermions is not much
  heavier than itself. In fact, if such a \neut{1} has a mass $\lesssim
  65$\gev, it can undergo annihilation via $h$ in the $s$-channel into a
  $b\bar{b}$ pair (see, e.g.,~\cite{Fowlie:2013oua, Bergeron:2013lya}). For even smaller
bino masses, the right amount of $s$-channel annihilation can occur through the
$Z$ boson. Evidently, the cross section for annihilation via $h$ or
$Z$ falls abruptly as the
difference between \mneut{1} and $m_h/2$ or $m_Z/2$, respectively,
increases. Below $\mneut{1} \lesssim 30$\gev\ it is thus generally impossible
to obtain sufficient \omegadm\ in
the MSSM due to the absence of a mediator with the correct mass.

The Next-to-Minimal Supersymmetric Standard Model
(NMSSM)~\cite{Fayet:1974pd,Ellis:1988er,Durand:1988rg,Drees:1988fc}
 contains a gauge singlet Higgs superfield, $\widehat{S}$, in addition
 to the two doublet superfields, $\widehat{H}_d$ and $\widehat{H}_u$,
of the MSSM.
This results in two additional neutral mass eigenstates in the Higgs
sector besides the three similar to the
MSSM ones. Thus, in total there are seven Higgs bosons in the
NMSSM; CP-even $H_{1,2,3}$, CP-odd $A_{1,2}$
and a charged pair, $H^\pm$. Naturally, the two new Higgs bosons
are dominated by the scalar component of $\widehat{S}$,
which implies that their couplings to the
fermions and gauge bosons of the Standard Model (SM) are typically
much smaller than those of the doublet-like Higgs bosons.
Their masses are thus generally very weakly
constrained by the Higgs boson data from the Large Electron Positron
(LEP) collider, the Tevatron or the LHC, and can be
as low as a GeV or so. Similarly, the neutralino sector of the NMSSM also contains
five mass eignestates, one more than in the MSSM. The fifth neutralino is
dominated by the fermion component of $\widehat{S}$, commonly referred to as the
singlino.

In the NMSSM, the role of the $H_{\rm obs}$ can be played by either
one of the $H_1$ and
$H_2$~\cite{Ellwanger:2011aa,King:2012is,Ellwanger:2012ke,Gherghetta:2012gb,Cao:2012fz}
or even by a combination/superposition of these two when both of them
lie near 125\gev~\cite{Gunion:2012gc,Gunion:2012he,King:2012tr}.
Evidently, when $H_{\rm obs}$ is the $H_1$, only $A_1$ can be lighter
than it. Conversely, when $H_{\rm obs}$
is the $H_2$, there exist the possibilities of either the $H_1$ alone
or both the $H_1$ and $A_1$ being lighter than it.
Such light $H_1$ or $A_1$ can act as mediators for
the annihilation of a \neut{1} as light as a GeV or so in order for it to generate
the correct \omegadm\ - a scenario precluded in the MSSM. 
Note that the electroweak (EW) baryogenesis may also be accommodated
within this framework when the \neut{1} mass is below 
10~GeV~\cite{Carena:2011jy,Bi:2015qva}.
Additionally, a light NMSSM \neut{1} can explain the galactic centre $\gamma$-ray
excess~\cite{Hooper:2010mq,Hooper:2011ti} observed by the Fermi Large
Area Telesope in the presence of a light $A_1$~\cite{Cheung:2014lqa,Cao:2014efa,
Feng:2014vea,Bi:2015qva}.\footnote{In the case of an $H_1$, the
$s$-channel annihilation is $p$-wave suppressed and hence generally
negligible. Furthermore, due to the strong correlation between the
masses of $H_1$ and $H_2$, as will be shown later, it is very
difficult, although possible, to obtain an $H_1$ as light as a GeV or so
while requiring $H_2$ to have a mass near 125\gev.} If the \neut{1} 
is even lighter, it has
been shown~\cite{Kozaczuk:2013spa,Cao:2013mqa} to explain the CDMS-II
event excess near 10\gev~\cite{Agnese:2013rvf}. However,
further experimental evidence is necessary
to confirm whether any of these events are indeed caused by the annihilation
of the DM and the jury is still out on its correct mass.
While the direct-detection experiments such as 
DAMA/LIBRA~\cite{Bernabei:2010mq}, 
CoGeNT~\cite{Aalseth:2010vx,Aalseth:2011wp},
CRESST-II~\cite{Angloher:2011uu}, XENON100~\cite{Aprile:2012nq} and
 LUX~\cite{Akerib:2013tjd} can cover a wide range of the DM mass, it
 is extremely unlikely for them to reach below $\sim 2$\gev~\cite{Draper:2010ew}. 
 Similarly, while the IceCube neutrino observatory also shows some
 promise for the indirect detection of the NMSSM
 DM~\cite{Enberg:2015qwa}, the sensitivity for a mass below 10~GeV is very poor.

The Run-II of LHC can prove very crucial in this regard, owing to its
potential to probe a very light DM, produced in association
with the $Z$ boson or the $A_1$ in the decays of the heavier
neutralinos. In some recent
studies~\cite{Das:2012rr,Ellwanger:2013rsa,Kim:2014noa,Han:2014sya,Dutta:2014hma}
it has been shown that significant regions of the NMSSM parameter
space can be covered  by the $Z$-associated production of the \neut{1}
at the LHC Run-II, with the $Z$ decaying
into two leptons. The $\neut{2,3} \to A_1 + \neut{1}$ channel has also
shown some promise~\cite{Cerdeno:2013qta} even for the 8\tev\ LHC,
with the mass of the $A_1$
lying above the $\tau^+\tau^-$ production threshold. These search 
channels can be complemented by those where the light
$A_1$ is not accompanied by the DM. As long as the relic abundance
is satisfied for a given SUSY point with a very light DM,
observation of the corresponding $A_1$, which would be a likely 
mediator for its annihilation, could serve as a pointer.
Such channels include, most importantly, decays of the \hobs\ into
$A_1 + A_1$ or $A_1 + Z$~\cite{Ellwanger:2003jt,*Forshaw:2007ra,
*Almarashi:2010jm,*Almarashi:2011hj,*Almarashi:2011qq,*Ellwanger:2013ova,*Cao:2013gba}.
At the Run-I of LHC, light pseudoscalars decaying in the
$\mu^+\mu^-$ channel have already been probed, when they are
produced either singly in $pp$ collisions~\cite{Chatrchyan:2012am}
or in pairs from the decays of a 125\gev\ Higgs
boson~\cite{Chatrchyan:2012cg}. These searches thus strongly constrain the
parameter space of the NMSSM. Prospects for the observability of $A_1$
 at the 14\tev\ LHC have been analyzed in detail recently
in~\cite{Bomark:2014gya,Curtin:2014pda,Bomark:2015fga}, with the
measurement of the mass of \hobs\ providing an additional handle
in the kinematic selection of the events. The channels
investigated in all these analyses, however, show sufficient
sensitivity only for an $A_1$ heavier than 5\gev.

In this study we focus on the production of a very light,
$\mathcal{O}(1)$\gev, \neut{1} via decays of the heavier, higgsino-like,
neutralinos of the NMSSM. We perform detailed scans of the NMSSM parameter space
to highlight its regions where such a \neut{1}, consistent with the observed
relic abundance of the universe, can be obtained in the presence of a
SM-like $H_2$. We discuss some important characteristics of
these parameter regions and of the \neut{1}. We note, in particular,
that the \neut{1} is always accompanied by a
pseudoscalar with a mass nearly twice its own. We then carry out
comprehensive detector-level analyses of the $Z$- as well as
 $A_1$-associated productions of the \neut{1}. For the $A_1 + \neut{1}$
production channel, we adopt an unconventional method for
reconstructing the pair of highly collinear muons that result
from the subsequent decay of the $A_1$. We find that by using our
signal reconstruction method the $A_1 + \neut{1}$ channel can have a better
signal strength at the 14\tev\ LHC compared to the
$Z + \neut{1}$ channel for certain specific parameter configurations.
Therefore, the $A_1 + \neut{1}$ channel can serve as a
crucial probe of new physics, complementing well the $Z + \neut{1}$
channel for very a low-mass singlino-dominated DM.

The article is organized as follows. In section~\ref{sec:model}, we will
briefly discuss the NMSSM and its singlet sector. In
section~\ref{sec:method} we will present some details of our findings from
the numerical scans of the NMSSM parameter space. In
section~\ref{sec:events} we will explain our signal-to-background
analyses and in section~\ref{sec:results} we will show their results. 
We will present our conclusions in section~\ref{sec:concl}.

\section{\label{sec:model} The singlet sector of the NMSSM}

The scale-invariant superpotential of the NMSSM (see,
e.g.,~\cite{Ellwanger:2009dp,Maniatis:2009re} for reviews) is written as
\begin{equation}
\label{eq:superpot}
W_{\rm NMSSM}\ =\ {\rm MSSM\;Yukawa\;terms} \: +\:
\lambda \widehat{S} \widehat{H}_u \widehat{H}_d \: + \:
\frac{\kappa}{3}\ \widehat{S}^3\,,
\end{equation}
where \lam\ and \kap\ are dimensionless Yukawa couplings. The above
superpotential observes a discrete $Z_3$ symmetry, which forbids the $\mu
\widehat{H}_u \widehat{H}_d$ term present in the MSSM superpotential
and at the same time breaks the dangerous Peccei-Quinn (PQ) symmetry~\cite{Peccei:1977hh,*Peccei:1977ur}.
Here, an effective $\mu$-term is generated upon spontaneous
symmetry breaking, when $\widehat{S}$ develops a vacuum
expectation value (VEV), $s\equiv\langle \widehat{S}\rangle$, so that
$\mueff=\lam s$. The soft SUSY-breaking terms in the scalar Higgs
sector are then given by
\begin{eqnarray}
V_{\rm{soft}}=m_{\hu}^2|\hu|^2+m_{\hd}^2|\hd|^2+m_{S}^2|S|^2+\left(\lam\alam
  S \hu \hd+\frac{1}{3}\kap\akap
  S^3+\textrm{h.c.}\right)\,.
\label{eq:vsoft}
\end{eqnarray}
Through the minimization conditions of the complete tree-level Higgs
potential, the soft masses $m_{\hu}$, $m_{\hd}$ and $m_S$ can be
traded for the respective VEVs, $v_u$, $v_d$ and $s$, of the corresponding Higgs
fields.

The neutral scalar and pseudoscalar Higgs mass matrices are obtained from
the Higgs potential evaluated at the vacuum. Diagonalization
of these matrices yields the mass expressions for the five neutral Higgs
bosons. The tree-level masses of the two lightest CP-even Higgs bosons, which
are of our interest here, can be approximated, for moderate-to-large
\tanb\ ($\equiv v_u/v_d$) and EW-scale dimensionful parameters, 
by~\cite{Miller:2003ay}
\begin{eqnarray}
\label{eq:mh12form}
m_{H_{1,2}}^2\approx&&\frac{1}{2}\Big\{m_Z^2+4(\kap s)^2+\kap s
  \akap\Big. \nonumber \\ 
&&\Big. \mp\sqrt{\left[m_Z^2-4(\kap s)^2-\kap s
      \akap\right]^2+4\lam^2 v^2\left[2\lam s-\left(\alam+\kap
        s\right)\sin 2\beta\right]^2}\Big\}\,,
\end{eqnarray}
where $v \equiv \sqrt{v_u^2 + v_d^2} \simeq 174$\gev.
Thus there is a strong correlation between the masses of $H_1$ and
$H_2$, implying that requiring one of these to lie near
125\,GeV constrains the other also.
Similarly, the tree-level mass of the singlet-like pseudoscalar
can be given, assuming negligible singlet-doublet mixing, by the
approximate expression
\begin{equation}
\label{eq:ma1}
m_{A_1}^2 \simeq \lam(\alam+4\kap s)\frac{v^2 \sin 2\beta}{2s}-3\kappa
sA_\kappa\,.
\end{equation}

The neutralino mass matrix in the NMSSM is written as
\begin{eqnarray}
\hspace*{0.2cm}
{\cal M}_{\widetilde{\chi}^0} =
\begin{pmatrix}
M_1 	& 0 	        &  -m_W \tan\theta_W \cos\beta      & m_W \tan\theta_W \sin\beta & 0  	\\
 0	& M_2 	& m_W \cos\beta                              &-m_W   \sin\beta               & 0  	\\
 -m_W \tan\theta_W \cos\beta  	&   m_W \cos\beta  	& 0				& -\mueff				& -\lambda v_u \\
 m_W \tan\theta_W \sin\beta	&  -m_W   \sin\beta   	& 	-\mueff			& 0				& -\lambda v_d \\
 0  	&  0	& 		-\lambda v_u		& 	 -\lambda v_d 			& 2\kappa s
\end{pmatrix}\,, \nonumber\\
\label{eq:massmatrix}
\end{eqnarray}
with $m_W$ being the mass of the $W$ boson and $\theta_W$ being the
weak mixing angle.
This symmetric mass matrix can be diagonalized by a unitary
matrix $N$ to obtain the five neutralino states, \neut{1-5}, ordered
by their masses, implying that the \neut{1} is the LSP. This mass eigenstate is
given by the linear combination
 \begin{eqnarray}
 \neut{1} = N_{11} \widetilde B^0 + N_{12} \widetilde W_3^0 + N_{13} \widetilde
 H_d^0 + N_{14} \widetilde H_u^0 + N_{15} \widetilde S^0.
 \end{eqnarray}
We define the gaugino fraction in \neut{1} as
$Z_g=|N_{11}|^2+|N_{12}|^2$, the higgsino
fraction as $Z_h=|N_{13}|^2+|N_{14}|^2$
and the singlino fraction as $Z_s=|N_{15}|^2$. The composition of \neut{1} thus
depends on the values of the various model parameters appearing in the
above mass matrix.

In the MSSM, the fifth row and column, corresponding to the singlino
and hence containing the $\lambda$-dependent terms, do not
exist. There, the smaller the $\mu$ compared to the $\min[M_1,M_2]$, the
larger the higgsino fraction in \neut{1}.
Since the mass of the higgsino-like chargino, which is the lighter
\chgo{1} in this case, is also
proportional to $\mu$, the value of this parameter is bounded from
below to around 100\,GeV by the non-observation of a chargino at
the large electron positron (LEP) collider. On the other hand, in
order to avoid excessive fine-tuning of the model parameters
for obtaining the correct Higgs boson mass, $\mu$ ought not to be too
large. However, a purely higgsino LSP does
not give the correct thermal relic density for masses below $\sim 1$\,TeV,
as the annihilation cross section becomes too large.
Therefore, the only way for the \neut{1} in the
MSSM to give good relic abundance with a mass $\mathcal{O}(10)$\gev\
is to have a sizable bino component~\cite{Bergeron:2013lya,Han:2014nba}.

In the NMSSM, the presence of the singlino leads to some additional
possibilities in the context of DM phenomenology. In the limit
$\mueff \ll \min[M_1,M_2]$, the term $[\mathcal{M}_{\widetilde{\chi}^0}]_{55} = 2\kap s =
2\frac{\kap \mueff}{\lam}$ in eq.~(\ref{eq:massmatrix}) implies that
the LSP is singlino-dominated for $2\kap / \lam < 1$. The singlino
fraction in \neut{1} can be increased by reducing \kap. One could
alternatively increase \lam, while keeping \mueff\ fixed, to reduce the
 size of the $[\mathcal{M}_{\widetilde{\chi}^0}]_{55}$ term, but this
 also enlarges the sizes of the
off-diagonal mixing terms. Eq.~(\ref{eq:ma1}) shows that the
mass of $A_1$ also scales with $\kap s$. Thus a light singlino-like
\neut{1} can be naturally accompanied by a light $A_1$. Note that
the strong correlation, eq.~(\ref{eq:mh12form}), between the $H_2$
mass, which is required to be around 125\gev, and the $H_1$ mass
generally prevents the latter from taking very low values, although it also
scales mainly with $\kap s$. Importantly though, even if $H_1$ has a
mass close to 2\neut{1}, the \neut{1} annihilation via $s$-channel $H_1$
is $p$-wave suppressed, which would make it extremely difficult to
generate the correct thermal relic abundance. We therefore focus
only on a light $A_1$ here, which can result in the singlino-like
\neut{1} yielding the correct relic density even with its mass around a
GeV. 

\begin{figure}[tbp]
\centering
\subfloat[]{%
\includegraphics[scale=0.7]{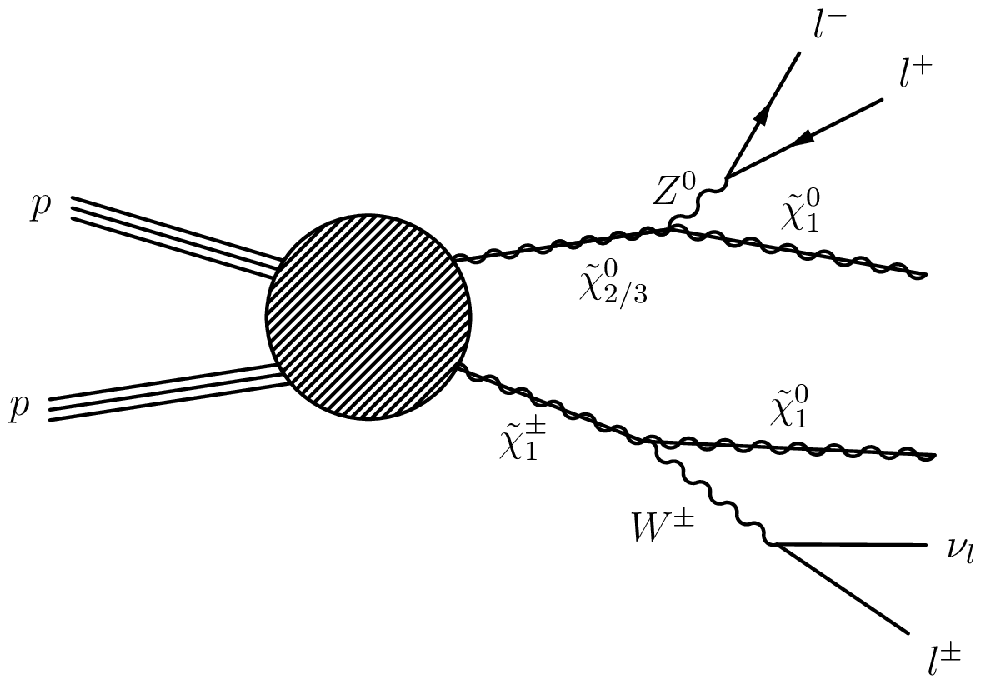}
}%
\hspace{1.0cm}%
\subfloat[]{%
\includegraphics[scale=0.7]{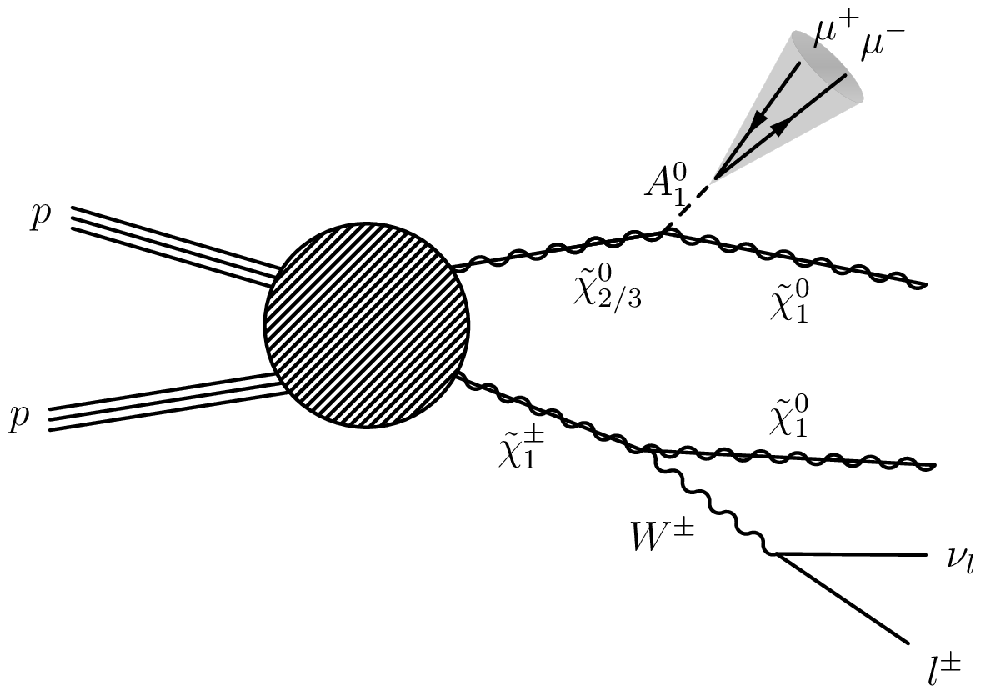}%
}%
\caption[]{Production process for the \neut{2,3}\chgo{1} pair, with the
  \neut{2,3} decaying via (a) $Z + \neut{1}$ and (b) $A_1 +
  \neut{1}$.}
\label{fig:process}
\end{figure}

As noted in the Introduction, the light \neut{1} can be produced in the
decays of the heavier neutralinos. In particular, the \neut{2} and
\neut{3} as well as the \chgo{1} in this scenario with a light singlino-like DM
are predominantly higgsinos~\cite{Cerdeno:2013qta}.
Their main production channel is $pp \to \neut{2,3} + \chgo{1}$, which is
followed by the decays $\neut{2,3}\to Z
/ A_1 + \neut{1} \to \ell^+\ell^- + \met$ and $\chgo{1}\to
W^\pm + \neut{1} \to \ell^\pm + \met$, where \met\ implies missing
transverse energy. The complete processes are shown in
Fig.~\ref{fig:process}, (a) and (b). The $Z + \neut{1}$
decay channel of the \neut{2,3} is by far the dominant one, while
$\neut{2,3} \to A_1 + \neut{1}$ is generally suppressed.
The $Z + W + \met$ production, which results in trileptonic
final states, is therefore the preferred search channel for the
higgsinos, also since it has minimal dependence on \mneut{1}.
Dedicated searches carried out by both the CMS and ATLAS
collaborations~\cite{atlas-3l,Aad:2014nua,Khachatryan:2014qwa}
 have either already put strong constraints on
significant regions of the NMSSM parameter space
or are likely to cover large portions of it, where the \neut{1} can be
as light as a GeV, at the LHC Run-II.

It is, however, important to note that, for sufficiently
large values of the parameter \lam, the branching ratio (BR) for the
$\neut{2,3} \to A_1 + \neut{1}$ decay channel, while still
subdominant, can become sizable.
For $A_1$ masses of interest here, i.e., below the $\tau^+\tau^-$
threshold, the decay mode with the largest BR is $A_1 \to c\bar{c}$.
The main leptonic decay channel, $A_1 \to \mu^+ \mu^-$, is highly
 subdominant, with its BR
never exceeding 9\%, as we shall see below. This value is still
larger than twice that of the BR$(Z \to \mu^+ \mu^-)$ ($\simeq 0.034$).
But one challenge that arises in this context is the reconstruction of the very
light $A_1$ from two highly collinear muons it decays into and the
isolation of this signal from the background. As long as this complication can be
overcome, the two observations above imply that the $A_1 + \neut{1}$ search channel
can possibly complement the $Z\neut{1}$ channel for certain specific
NMSSM parameter space points. In the following sections we will
analyze such parameter combinations and also introduce our
 method for reconstructing the very light $A_1$ from two collinear
 muons. We should point out here that while in principle the same method can
 alternatively be used to probe these $A_1$'s in the
decays of the heavy CP-even Higgses, our requirement of the presence
of \met\ in the final state, as an indication of light DM,
renders these channels irrelevant here.



\section{\label{sec:method} Parameter space scan and constraints}

Due to the presence of the additional singlet superfield, the NMSSM
contains several new parameters besides the 115 or so of the MSSM
at the EW scale. However,
assuming the sfermion mass matrices and the scalar trilinear coupling
matrices to be diagonal reduces the parameter space of the model considerably.
In this study, since our main focus is the higgsino-singlino sector, we
further assume the following universality conditions.
\begin{gather}
M_0 \equiv M_{Q_{1,2,3}} = M_{U_{1,2,3}} = M_{D_{1,2,3}} =
M_{L_{1,2,3}} = M_{E_{1,2,3}}\,, \nonumber \\
M_{1/2} \equiv 2M_1 = M_2 =  \frac{1}{3} M_3\,, \\
A_0 \equiv A_t = A_b = A_{\tau}\,, \nonumber
\end{gather}
 where $M_{Q_{1,2,3}},\,M_{U_{1,2,3}},\,M_{D_{1,2,3}}
 ,\,M_{L_{1,2,3}}$ and $M_{E_{1,2,3}}$ are the soft masses of the
 sfermions, $M_{1,2,3}$ those of the gauginos and $A_{t,b,\tau}$ the
trilinear Higgs-sfermion couplings. Along with $M_0$, $M_{1/2}$ and
$A_0$, the model then contains \lam, \kap, \mueff, \tanb, \alam\ and
\akap\ as the only free parameters, which are input at the
SUSY-breaking scale, $M_{\rm SUSY} = \sqrt{m_{\tilde{t}_1}
  m_{\tilde{t}_2}}$, with $m_{\tilde{t}_{1,2}}$ being the physical
    masses of the two stops.

We performed a scan of the above mentioned nine parameters using the
nested sampling package
\MTN-v2.18~\cite{Feroz:2008xx}, which was interfaced with
\NMT-v4.5.0~\cite{Ellwanger:2004xm,Ellwanger:2005dv,Das:2011dg,NMSSMTools}
for calculation of the SUSY
mass spectrum and BRs for each model point
sampled. The scanned ranges of the parameters are given in
table~\ref{tab:params} and were chosen, based loosely
on the findings of~\cite{Cerdeno:2013qta}, to maximally yield
points with $\mneut{1} \lesssim 2$\gev\ as well as the $H_2$ having a
mass near the one measured for the \hobs\ at the LHC. The $H_2$
was additionally required to have SM-like signal rates. The signal
rate, obtained from \NMT\ as an output, is defined, for a given decay 
channel $X$, as
\begin{equation}
R^X \equiv  \frac{\sigma(gg\rightarrow H_2)\times {\rm BR}(H_2\rightarrow
  X)}{\sigma(gg\rightarrow h_{\rm SM})\times {\rm BR}(h_{\rm SM} \rightarrow X)}\,,
\end{equation}
where ${h_{\rm SM}}$ is the SM Higgs boson with a mass equal to $m_{H_2}$. 

\begin{table}[tbp]
\begin{center}
\begin{tabular}{|c|c|}
\hline
 Parameter & Scanned range  \\
\hline
 \mzero\,(GeV) & 500 -- 2000\\
\mhalf\,(GeV) & 300 -- 1000\\
\azero\,(GeV)  & 0 -- 4000  \\
\mueff \,(GeV) & 100 -- 300 \\
\tanb & 6 -- 25  \\
\lam  & 0.01 -- 0.4   \\
\kap  & 10$^{-5}$ -- 10$^{-1}$ \\
\alam\,(GeV) & 0 -- 5000 \\
\akap\,(GeV) & $-100$ -- 0  \\
\hline
\end{tabular}
\caption{Ranges of the NMSSM input parameters scanned for obtaining
 \neut{1} with a mass below 2\gev.}
\label{tab:params}
\end{center}
\end{table}

A very light $A_1$ is subject to constraints from direct 
collider searches as well as from flavor physics~\cite{Dolan:2014ska}. 
The program \NMT\ intrinsically takes into account the exclusion
limits on pseudoscalars from LEP and BaBar as well as from the 4$\mu$ 
searches at the CMS~\cite{Chatrchyan:2012cg}. Additionally, It
ensures that the LEP limit on the \chgo{1} mass and the
perturbativity constraints on the various Higgs boson couplings are
satisfied and that the \neut{1} is the LSP.
We further required each scanned point to satisfy the following constraints:
\begin{itemize}
\item $2.63 \times 10^{-4} \leq \brbxsgamma \leq 4.23 \times 10^{-4}$,
\item $0.71 \times 10^{-4} \leq \brbutaunu \leq 2.57 \times 10^{-4}$,
\item  $1.3 \times 10^{-9} \leq \brbsmumu \leq 4.5 \times 10^{-9}$,
\item $0.107 \leq \Omega_{\neut{1}} h^2 \leq 0.131$,
\item $122\gev \leq m_{H_2} \leq 128\gev$.
\end{itemize}
The $b$-physics observables above were calculated using the package
\SI-v3.4~\cite{superiso} and the relic density using the
\MO-v4.1.5~\cite{Belanger:2014vza} package. The allowed range of
$\Omega_{\neut{1}}$ assumes a $\pm 10\%$ theoretical error
around the central value of 0.119 measured by the PLANCK
telescope~\cite{Planck:2015xua}. Similarly an error of $\pm 3\gev$ is
allowed in the theoretical estimation of the $H_2$ mass, given the
\hobs\ mass measurement of 125\gev\ at the LHC. Consistency with 
the LEP and LHC exclusion limits on the non-SM-like Higgs bosons, 
including the $A_1$, was further ensured by testing each point that passed
 the above constraints with the program
\HiB-v4.2.0~\cite{Bechtle:2008jh,Bechtle:2011sb,Bechtle:2013gu,Bechtle:2013wla}.

\begin{figure}[tbp]
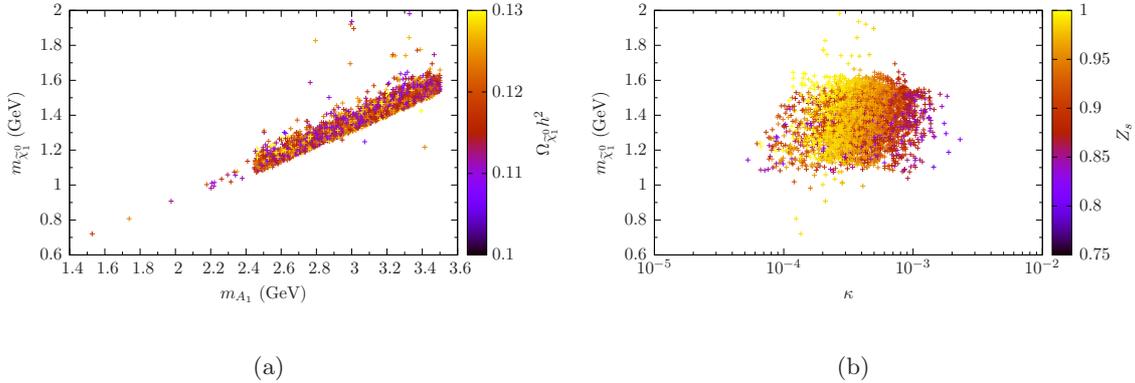

\hspace{-0.3cm}%
\centering
\subfloat[]{%
\resizebox{0.5\textwidth}{!}{\input{ma1_mchi1_RD.tex}}
}%
\subfloat[]{%
\resizebox{0.5\textwidth}{!}{\input{mchi1_kap_Zs.tex}}
}%
\caption[]{(a) The distribution of the \neut{1} mass vs. that of the $A_1$
mass, with the heat map corresponding to the relic abundance. (b) The dependence
of \mneut{1} on the parameter \kap. The heat map shows the singlino
fraction in the \neut{1}.}
\label{fig:ma1mchi1}
\end{figure}

\begin{figure}[tbp]
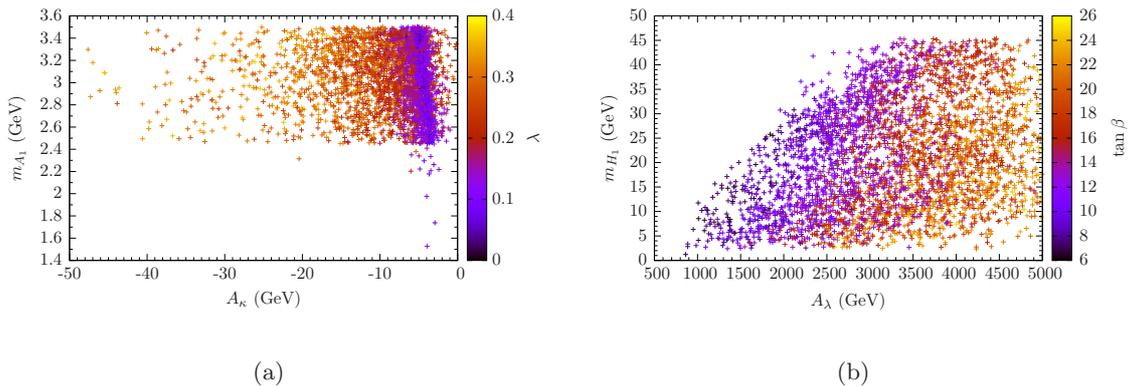

\hspace{-0.3cm}%
\centering
\subfloat[]{%
\resizebox{0.5\textwidth}{!}{\input{ma1_akap_lam.tex}}
}%
\subfloat[]{%
\resizebox{0.5\textwidth}{!}{\input{mh1_alam_tb.tex}}
}%
\caption[]{(a) The mass of $A_1$ as a function of the parameters \akap\
  and \lam. (b) The mass of $H_1$ vs. the distributions of the parameters
  \alam\ and \tanb.}
\label{fig:compo}
\end{figure}

In Fig.~\ref{fig:ma1mchi1}(a) we show the ranges of the $A_1$ and \neut{1}
masses obtained for the good points from our scan. The heat map
corresponds to the \neut{1} relic density. One can see a strong
correlation between \mneut{1} and $m_{A_1}$, with the former almost
always being half of the latter in order for the resonant
annihilation of the \neut{1} via the $A_1$. The tiny \neut{1} masses are
a result of vanishing \kap\, as seen in Fig.~\ref{fig:ma1mchi1}(b),
implying an almost PQ-symmetric model. The heat map in the figure
shows the distribution of the singlino fraction in the \neut{1} which
increases as \kap\ decreases and is always larger than 0.75. A large
singlino component implies a small higgsino fraction,
which is necessary to prevent an under abundance of the \neut{1} from
too much annihilation.
Fig.~\ref{fig:compo}(a) illustrates that the $A_1$ is restricted to a lower
mass, needed for satisfying the relic density constraint, by
adjusting \akap\ to a narrow range of low negative values. This
is in agreement with eq.~(\ref{eq:ma1}), along with the fact
that \lam, illustrated by the heat map in the figure, also
generally tends to be small. In Fig.~\ref{fig:compo}(b) the
distributions of the parameters \alam\ and \mueff\ are shown against the
$H_1$ mass range obtained in the scan. We see that when both \alam\
and \tanb\ are small $m_{H_1}$ is low, while its maximum value, $\sim
45\gev$, is obtained for the largest allowed values of \alam, with
$\tanb \gtrsim 10$.

\begin{figure}[tbp]
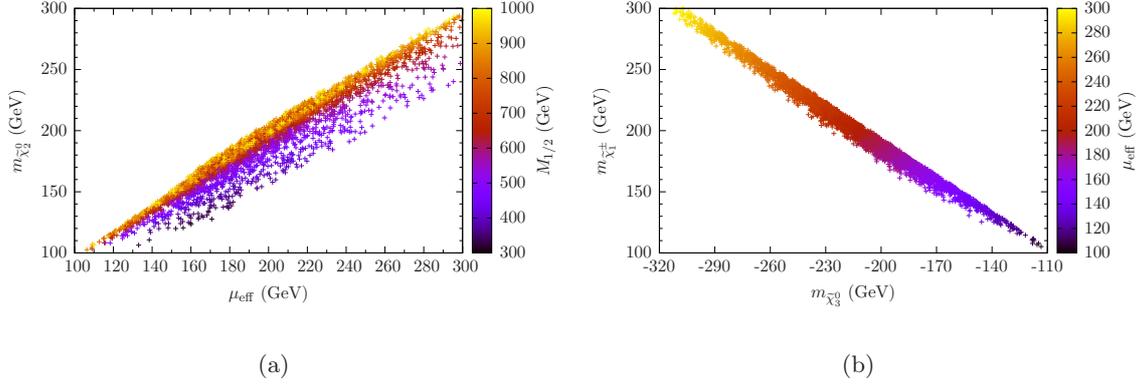

\hspace{-0.3cm}%
\centering
\subfloat[]{%
\resizebox{0.5\textwidth}{!}{\input{mchi2_mu_m12.tex}}
}%
\subfloat[]{%
\resizebox{0.5\textwidth}{!}{\input{mchi3_mchg1_mu.tex}}
}%
\caption[]{(a) The \neut{2} mass as a function of the parameter
  \mueff. The color map shows the dependence on \mhalf.
(b) The distribution of the \neut{3} and \chgo{1}  masses vs. \mueff,
shown by the heat map.}
\label{fig:mchi2mchi3}
\end{figure}

Fig.~\ref{fig:mchi2mchi3}(a) shows that the mass of the \neut{2} is
almost equal to the parameter \mueff\ as long the parameter \mhalf,
given by the color map, approaches its maximum allowed value. As the
splitting between \mhalf\ and \mueff\ decreases, the gaugino-higgsino
mixing increases, which results in somewhat lowering \mneut{2} relative to
\mueff. Fig.~\ref{fig:mchi2mchi3}(b) similarly shows that, since the \neut{3}
and the \chgo{1} are higgsino-like as well, their masses are also
proportional to \mueff. Moreover, the \chgo{1} is
always almost mass-degenerate with \neut{2}, while \neut{3} is
generally heavier than both of them, but only slightly
so.\footnote{Due to the nature of the
  neutralino mass matrix, given in eq.~(\ref{eq:massmatrix}), both
  positive and negative values of the \neut{3} mass are possible,
although we show here only
  negative valued solutions. Positivity of \mneut{3} can be ensured 
{\it a priori} by a phase transformation.}
Fig.~\ref{fig:BRschi2chi3}(a) illustrates that
the BR($\neut{2} \to A_1 \neut{1}$) increases with \lam, shown by the
heat map. This is because, for vanishing singlet-doublet mixing in 
the pseudoscalar mass matrix as well as singlino-higgsino mixing in 
the neutralino mass matrix, the coupling between the $A_1$, $\neut{1}$ 
and $\neut{2}$, given by eq.~(A.14) of~\cite{Ellwanger:2009dp}, can 
be approximated by the simple relation
\begin{equation}
g_{A_1\neut{1}\neut{2}} \approx -\frac{i}{20}(\lam-2\sqrt{2}\kap)\,.
\end{equation}
Still, the BR($\neut{2} \to A_1 \neut{1}$) exceeds 10\% only for a few 
points, reaching
as high as about 30\% for a couple of them. On the other hand,
the BR($\neut{2} \to Z \neut{1}$), given on the $x$-axis, drops quite
sharply with increasing \lam\ and can reach as low as about 30\%.
This is due mainly to the fact that the BR($\neut{2} \to
H_2 \neut{1}$) (not shown here) also rises, much more abruptly than
 the BR($\neut{2}\to A_1 \neut{1}$), as \lam\
increases. Fig.~\ref{fig:BRschi2chi3}(b) shows that the BR($\neut{3}
\to A_1 \neut{1}$) never exceeds 9\% while the BR($\neut{3}
\to Z \neut{1}$) never falls below 50\%, implying a relatively weaker
dependence on \lam. The heat map in the figure shows that the
 maximum BR($A_1 \to \mu^+\mu^-$) achievable is 9\%. Note
 that the BR($\chgo{1} \to W^{\pm} \neut{1}$) is unity for all the
 points shown in these figures.

\begin{figure}[tbp]
\hspace{-0.3cm}%
\centering
\subfloat[]{%
\resizebox{0.5\textwidth}{!}{\input{BRchi2a1_BRchi2z_lam.tex}}
}%
\subfloat[]{%
\resizebox{0.5\textwidth}{!}{\input{BRchi3a1_BRchi3z_mu.tex}}
}%
\caption[]{(a) The BR($\neut{2} \to Z \neut{1}$) and the BR($\neut{2} \to A_1
  \neut{1}$) as functions of the parameter \lam. (b)
The BR($\neut{3} \to Z \neut{1}$) vs. the BR($\neut{3} \to A_1
\neut{1}$), with the heat map corresponding to the BR($A_1 \to \mu^+\mu^-$).}
\label{fig:BRschi2chi3}
\end{figure}

We point out here that we also carried out a test scan of the NMSSM with
partial universality at the grand unification (GUT) scale. In
this model the unified scalar mass
parameter, $m_0$, the universal gaugino mass, $m_{1/2}$, the
universal Higgs-sfermion trilinear couplings, $a_0$, as well as the
paramaters $a_\lam$ and $a_\kap$ are input at the GUT scale. The EW scale
values of the individual soft scalar and gaugino masses and of all the
Higgs trilinear couplings are obtained by the running of these parameters
using the renormalization group equations. The other Higgs sector
parameters, \lam, \kap, \mueff\ and \tanb\ are input at $M_{\rm
  SUSY}$. But, owing particularly to the GUT scale definition of the parameters
$a_\lam$ and $a_\kap$, severe fine-tuning is necessary in this model
for the scan to find the right combinations of these parameters
at $M_{\rm SUSY}$ that yield both $H_2$ and $A_1$ with the desired masses.

As noted above in Fig.~\ref{fig:compo}(a), \akap\ at $M_{\rm SUSY}$ is
restricted to a narrow range of values, which would
imply an even smaller set of its possible values at the GUT scale.
Furthermore, the SUSY-preserving parameter \kap\ is also very small
due to our requirement of $\mneut{1} \sim 1.5$\,GeV, though the
approximate PQ symmetry this results in still avoids the cosmological
constraints on the PQ axion~\cite{Lebedev:2009ag}.
Keeping \mueff\ around the EW scale, \kap\ is typically smaller than
$0.01\lam$. Despite all these limiting conditions, some
 points with the right masses of the \neut{1} and $A_1$ were indeed found.
However, they did not cover, for example, the wide range of the
BR($\neut{2} \to A_1 \neut{1}$) seen in fig.~\ref{fig:BRschi2chi3}(a),
which stretches beyond 0.1. We therefore opted for the 9-parameter
EW-scale NMSSM for this study.


\section{\label{sec:events} Very light DM via higgsino decays
at the LHC}

For our signal-to-background analyses, we selected two benchmark points,
BP1 and BP2, out of the good points from the NMSSM parameter space scan.
BP1 is chosen such that the BR($\neut{2,3} \to Z \neut{1}$) is sufficiently
large, while BP2 has a relatively enhanced BR($\neut{2,3} \to A_1
\neut{1}$). The parton-level signal and background
events for these points were generated with
\MadG\_aMC@NLO~\cite{Alwall:2014hca} and passed to
\pythia\ 6.4~\cite{Sjostrand:2006za} for hadronization.

\subsection{The trilepton channel}

As noted earlier, the ATLAS and CMS experiments have separately
performed searches for trileptons 
($3\ell$)~\cite{atlas-3l,Aad:2014nua,Khachatryan:2014qwa}
resulting from the $\neut{2,3} \chgo{1}$ pair production.
In the ATLAS $3\ell$ search~\cite{atlas-3l}, which is the one we will
consider in the following, six signal regions (SRs) are defined
in terms of the invariant mass of two same flavor leptons with
opposite sign (SFOS), $m_{\rm SFOS}$. These regions also depend 
on the momentum, $p_{T(\ell_3)}$,  of the third lepton, $\ell_3$, that is left
after requiring two SFOS leptons to 
reconstruct $m_{\rm SFOS}$, \met\ and the transverse mass,
$M_T =\sqrt{2\,\met\,p_{T(\ell_3)}\,(1-\cos \Delta \phi_{\ell_3,\met})}$,
where $\Delta\phi_{\ell_3,\met}$ is the azimuthal angle
between \met\ and the $\ell_3$. The signal regions are divided into 
three `$Z$-enriched' ones,
SRZ\{a,b,c\}, where $m_{\rm SFOS}$ lies within 10\gev\ of $m_Z$, and three
`$Z$-depleted' ones, SRnoZ\{a,b,c\}, where $m_{\rm SFOS}$ lies outside
this mass window. Table~\ref{tab:trilep} further shows the selection
requirements for each of these six regions.

\begin{table}[tbp]
\small
\centering \begin{tabular}{|c|c|c|c|c|c|c|c|}
\hline  Selection   & SRnoZa         & SRnoZb      & SRnoZc   & SRZa  & SRZb & SRZc \\
\hline  $m_{\rm SFOS}$ & $< 60$ 	& $60-81.2$ 	&$< 81.2$ or
$>101.2$	& $81.2-101.2$ & $81.2-101.2$ 	& $81.2-101.2$  \\
\hline  \met\  &$ > 50$ 	& $> 75$ 	&$>75$ 	&$75-120$ 	&$75-120$ 	&$>120$  \\
\hline  $M_T $  &$-$ 	&$-$	&$>110$ 	&$<110$ 	&$>110$ 	&$>110$  \\
\hline  $p_{T(\ell_3)}$  &$>10$    &$>10$ 	&$>30$ 	&$>10$ 	&$>10$ 	&$>10$  \\
\hline  SR veto   & SRnoZc   & SRnoZc & $-$ & $-$ & $-$ & $-$ \\
\hline
\end{tabular}
\caption{Selection requirements for the six signal regions defined
  for the trilepton searches by the ATLAS collaboration.
All the dimensionful parameters in  rows $2-5$ are in units of GeV.}
\label{tab:trilep}
\end{table}

The irreducible backgrounds include di-boson, tri-boson and
$t\bar{t}W/Z$ productions, all of which can have three or
more leptons and \met\ in the final states. The $ZZ$ and $ZW^{\pm}$
backgrounds are by far dominant over the $t\bar{t}W/Z$ one~\cite{atlas-3l}.
Among the reducible backgrounds are included top
quarks produced singly or in pairs, $WW$ and $W$ or $Z$ bosons
produced in association with jets or photons. Among these the $t\bar{t}$
background is highly dominant.
For each BP, the cross section for the signal process was calculated
at next-to-leading order (NLO) using \Prosp-v2.1~\cite{prospino}.
We then first generated the event files corresponding to $\sqrt{s} = 8$\tev\
for the signal process and passed these to
the public package \cmate-v1.2.0~\cite{Drees:2013wra} for testing
against the current LHC limits from the trilepton searches. In \cmate\
the signal regions given in table~\ref{tab:trilep} have been
defined and the corresponding backgrounds from the
ATLAS experiment implemented. For testing a model point it
therefore calculates the signal efficiency for each region, after
ATLAS detector simulation with \DELP\ 3~\cite{deFavereau:2013fsa}.

After confirming that the given BP is not excluded by the available
data, we proceeded to the future $3\ell$ search at the 14\tev\ LHC.
We generated the signal event files for $\sqrt{s} = 14$\tev\ and
passed these to \cmate\ again. In this way we obtained the signal
efficiencies, after multiplying the NLO cross section with an assumed 
integrated luminosity, $\mathcal{L}$, of 300\,fb$^{-1}$ (i.e.,
design luminosity of the LHC) to get the number of signal events 
in each signal region.

As for the backgrounds, we only simulated the three
dominant ones, $ZZ$, $ZW^{\pm}$ and $tt$, for the 14\,TeV LHC. We used 
\MadG\ to 
generate the background events and passed these to \cmate\ to get the cut 
efficiencies for all the backgrounds in each signal region. After
multiplying the NLO cross section~\cite{Campbell:2011bn,Bonciani:1998vc} 
and the luminosity, we got the number of background events in each signal region.

\subsection{Collimated muons from an $A_1$}

Due to the smallness of the $A_1$ mass of our interest here, the
muons it decays into are highly collinear. In order to isolate such
muons, we employ the technique of clustering
them together into one object, \muco. This method, similar in
concept to the construction of a
`lepton-jet'~\cite{ArkaniHamed:2008qp,*ArkaniHamed:2008qn,*Baumgart:2009tn,*Katz:2009qq,*Cheung:2009su,*Falkowski:2010cm},
has recently been shown in~\cite{Han:2015lma} to be 
very effective for probing highly mass-degenerate higgsinos.

For using this method, the signal events generated for BP1 and BP2
were passed to \pythia\ 6.4 for hadronization and subsequently to 
\DELP\ 3 for jet-clustering via
the anti-$k_T$~\cite{Cacciari:2008gp} algorithm
using \Fastj-v3.0.6~\cite{Cacciari:2011ma}. The object
\muco\ is constructed as follows.
\begin{enumerate}
\item Require the transverse momentum, $p_T$, larger than 10\,GeV for each
muon in the signal. In addition, impose the cut
$m_{\mu\bar \mu}<5$\,GeV on the invariant mass of the muon pair.
\item Define $I_{\rm sum}$ as the scalar sum of the transverse momenta of
  all additional charged tracks, each with $p_T > 0.5$\,GeV, within a
  cone centered along the momentum vector of \muco\ and satisfying
  $\Delta R = 0.4$. Impose $I_{\rm sum} < 3 $\,GeV.
\end{enumerate}

The main backgrounds, containing two collinear muons along with a
third lepton and \met, include $W(\to \ell^{\pm}v)\gamma^*$ and $Z(\to
\ell^+\ell^-)\gamma^*$, wherein the \muco\ comes from the photon, and
$Wb\bar{b}$, when one of the $b$-jets produces the \muco.
The $Z\gamma^*$ background fakes the signal process when one of the 
two leptons escape undetected. Note that while the $t\bar{t}$ background
mentioned above for the $3\ell$ search is also relevant for this signal process, it
becomes negligible here owing to the
requirement of the two final state muons being highly collinear. In
order to isolate these backgrounds, we implement the following cuts.
\begin{enumerate}
\item Since the $A_1$ resulting from the higgsino decay is highly
  boosted, we expect the \muco\ from its subsequent decay to have a
  large $p_T$. We therefore require $p_{T(\muco)} > 50 $\,GeV. 
 We also require $p_{T(\ell_3)} > 20 $\,GeV.
\item In order to reduce the background a large \met\ is required, so
  we add the cut $\met > 50$\,GeV.
\item In the $W\gamma^*$ background the $M_T$ distribution has 
an end point around the $W$ boson mass, which leads us to impose $M_T > 80$\,GeV.
\item Our signal would appear as a narrow peak in the $m_{\muco}$
  distribution. Hence, we impose a narrow cut width,
  $5\times \sigma_{A_1}$, where $\sigma_{A_1}=0.26+0.0013m_{A_1}$, around $m_{A_1}$. 
This parametrization of $\sigma_{A_1}$ follows the mass resolution
study of the $J/\Psi$ resonance in~\cite{Chatrchyan:2011hr}. We also
remove the $J/\Psi$ resonance region ($3.0\gev < m_{\muco}< 3.2\gev$).
\end{enumerate}
To get a sufficient number of Monte Carlo events in the kinematic
regime of our interest, we require $m_{\mu\bar{\mu}} > 1.5$\,GeV
and $p_{T(\ell_3)} > 10$\,GeV at the parton level for the $W\gamma^*$ and
$Z\gamma^*$ backgrounds. For the $Wb\bar{b}$ background, we 
additionally require the $p_T$ of the $b$-jet to be larger than
30\,GeV.

\section{\label{sec:results} Results and discussion}

In table~\ref{tab:bench} are recorded some specifics of the two
BPs used for the signal-to-background analyses.
The consistency of each of the $H_2$
signal rates given in the last three rows of the table is to be checked against the
experimental quantity $\mu^X \equiv
\frac{\sigma(pp\to \hobs \to X)}{\sigma(pp\to h_{\rm SM} \to X)}$ for each
corresponding channel $X$. Note that this comparison assumes that
the inclusive $pp$ cross section for $H_2$ production can be
approximated by the dominant gluon-fusion production cross section.
Note also that, since the $WW$ and $ZZ$ decays of $H_2$ are
proportional to the same coupling, \NMT\ provides a unique value of
the signal rates for these two channels, which we denote by $R^{VV}$ in the table.
The most recent measurements of $\mu^X$ by the CMS (ATLAS)
collaboration(s) read~\cite{Khachatryan:2014jba,Aad:2014eha,ATLAS:2014aga,ATLAS-CONF-2014-009}
\begin{eqnarray}
\mu^{\gamma\gamma} &=& 1.13 \pm 0.24~(1.17 \pm 0.27)\,, \nonumber \\
\mu^{ZZ(WW)} &=& 1.0 \pm 0.29~(1.09^{+0.23}_{-0.21})\,, \\
\mu^{\tau\tau} &=& 0.91 \pm 0.28~(1.4^{+0.5}_{-0.4})\,. \nonumber
\end{eqnarray}

\begin{table}[th!]
\centering
\begin{tabular}{|c|c|c|}
\hline      	                   & BP1 & BP2 \\
\hline
\multicolumn{3}{|l|}{\it Model parameters} \\
\hline
\mzero\,(GeV) & 1951.1 & 1826.0 \\
\mhalf\,(GeV) & 892.24 & 929.2 \\
\azero\,(GeV)  & 2462.2 & 2626.2 \\
\mueff \,(GeV) & 191.34 & 164.52 \\
\tanb & 14.056 & 19.785 \\
\lam  & 0.0814 & 0.3102 \\
\kap  & 0.0002 & 0.0008 \\
\alam\,(GeV) & 4080.2 & 3596.7 \\
\akap\,(GeV) & $-3.6681$ & $-6.8466$  \\
\hline
\multicolumn{3}{|l|}{\it Masses} \\
\hline  \mneut{1} (GeV)  & 1.0025  & 1.4081    \\
\hline  \mneut{2} (GeV)   & 189.09  & 170.13   \\
\hline  \mneut{3} (GeV)   & $-201.67$  & $-182.27$    \\
\hline  \mchgo{1} (GeV)  & 194.97 & 167.72    \\
\hline  $m_{A_1}$ (GeV)    & 2.1776 & 2.9856     \\
\hline  $m_{H_2}$ (GeV)    & 124.12 & 125.79     \\
\hline
\multicolumn{3}{|l|}{\it Branching Ratios} \\
\hline  ~~$BR(\neut{2}\rightarrow Z \neut{1})$ & 0.634  & 0.603     \\
\hline  ~~$BR(\neut{2}\rightarrow A_1\neut{1})$ & 0.004  & 0.089   \\
\hline  ~~$BR(\neut{3}\rightarrow Z \neut{1})$ & 0.736  & 0.704      \\
\hline  ~~$BR(\neut{3}\rightarrow A_1\neut{1})$ & 0.004  & 0.081   \\
\hline  ~~$BR(A_1\rightarrow \mu^+\mu^-)$ & 0.039  & 0.087   \\
\hline
\multicolumn{3}{|l|}{\it $H_2$ signal rates} \\
\hline  $R^{\gamma\gamma}$  & 0.998 & 0.901   \\
\hline  $R^{VV}$ & 0.996  & 0.885    \\
\hline  $R^{\tau\tau}$ & 1.003 & 0.847   \\
\hline
\end{tabular}
\caption{Properties of the two benchmark points used for the
signal-to-background analyses.}
\label{tab:bench}
\end{table}

Using each of the two analysis methods described in the previous
section we calculated the number of signal events, $S$, and that
of background events, $B$, for each BP at the
LHC with $\sqrt{s} = 14$\tev\ and $\mathcal{L} = 300\,{\rm fb}^{-1}$. 
In table~\ref{tab:trilep-results} we provide the $S$ and $B$
corresponding to each of the six signal regions in the ATLAS $3\ell$
search. The total signal cross sections obtained are 24.3\,fb and 3.93\,fb 
for BP1 and BP2, respectively. In the $Z$-enriched region, $ZW^\pm$
production dominates the total background. In the $Z$-depleted
region, a comparable contribution is obtained from the $t\bar{t}$
background. One can notice in the table that for both the BPs, the
highest sensitivity is obtained in the signal region SRZc.

\begin{table}[th!]
\centering
\begin{tabular}{|c|c|c|c|c|c|c|c|}
\hline Background or signal & SRnoZa   & SRnoZb  & SRnoZc   & SRZa  & SRZb & SRZc \\
\hline  $ZZ$ events            &410 	&59 	&10 	&280 	&39 	&12  \\
\hline  $ZW^\pm$ events    &1391 	&595 	&71 	&6850 	&661 	&189  \\
\hline  $t\bar{t}$ events    &1715 	&401 	&62 	&272 	&178 	&19  \\
\hline  All background events & 3516 	&1055 	&143 	&7402 	&878 &220  \\
\hline  BP1 signal events        &12      &37       & 19 & 191 & 134 & 130 \\
\hline  BP2 signal events        &20      &46       & 18 & 270 & 144 & 96 \\
\hline
\end{tabular}
\caption{The number of background and signal events at the 14\,TeV LHC
  for $\mathcal{L}=300$\,fb$^{-1}$ in each of the signal regions of
  the ATLAS $3\ell$ search.}
\label{tab:trilep-results}
\end{table}

In the \muco\ channel, the cross sections for the signal and
background processes are given in table~\ref{tab:BP1-result} for the
BP1, along with the efficiency of the cuts and the effective cross
sections and numbers of events after implementing these cuts. One
notices that the $Wb\bar{b}$ is by far the largest background. It,
however, gets greatly reduced by the cuts, after
which $W\gamma^*$ takes over as the most dominant background. 
The corresponding quantities for the BP2  are listed in 
table~\ref{tab:BP2-result}. Due to the different masses of
the $A_1$ and \neut{1} obtained for the two BPs, different sets
of cuts need to be implemented for each of them.

\begin{table}[tbp]
\centering
\begin{tabular}{|c|c|c|c|c|c|c|c|}
\hline      & ~~BP1 ~~        & ~~$W\gamma^*$  ~~    & ~~$Z\gamma^*$~~   & ~~$Wb\bar{b}$  ~~ \\
\hline  Cross section  (fb)       &0.178 &246.9 & 10.0 & 3770.0\\
\hline  Cut efficiency  & 0.123 & $2.15\times 10^{-4}$ & $6\times 10^{-5}$ & $1\times 10^{-6}$ \\
\hline  Effective cross section (fb)     & 0.022 & 0.053 & 0.0006 & 0.003 \\
\hline No. of events & 6.6 & 15.9 & 0.18 & 0.9 \\
\hline
\end{tabular}
\caption{The backgrounds and the signal for the BP1 in the \muco\ search channel at the 14\,TeV LHC for $\mathcal{L}=300$\,fb$^{-1}$.}
\label{tab:BP1-result}
\end{table}

We quantify the strengths of the two analyses in terms of
$S/B$ for comparing them against each
other. This quantity is listed in table~\ref{tab:StoB}
for the two BPs in each of the search channel. The
$3\ell$ analysis gives a slightly higher $S/B$ for the BP1 compared to
that for the BP2. On the other hand, 
$S/B$ for the BP2 in the $\muco$ analysis is considerably larger 
than the $S/B$ obtained for the BP1 in each of the two analyses. This is due,
evidently, to the sizable BR($\neut{2} \to A_1
\neut{1}$) and BR($A_1\to \mu^+\mu^-$), as
noted in table~\ref{tab:bench}. In addition, the cut efficiency for
the signal is much higher while that for the $W\gamma^*$ background
is much lower in the case of the BP1. 

For a more realistic analysis of the prospects of a signal process
though, the statistical and systematic uncertainties in it also need to 
be taken into account. Hence, for each BP we have also provided in 
table~\ref{tab:StoB} the statistical
significance, given by the approximate formula,
\begin{eqnarray}
\mathcal{Z} \equiv \frac{S}{\sqrt{B + (\varepsilon B)^2}}\,,
\label{eq:significance}
\end{eqnarray}
where the systematic uncertainty is given by the fraction
$\varepsilon$ of the background. From the ATLAS $3\ell$ 
search~\cite{atlas-3l}, we note that
the systematic uncertainty is 21\% for the SRZc signal region, where the
highest sensitivity is achieved, as seen above. We
expect this number not to vary considerably at the 14\tev\ LHC and hence use
$\varepsilon = 0.21$ in our estimation of $\mathcal{Z}$ for
the $3\ell$ channel. For the $W+\muco+\met$ channel, since there is
no experimental analysis available, the systematic uncertainty has to be
estimated roughly. There are two major sources of this
uncertainty: the reconstruction of the \muco, in which
case it is around 5\%~\cite{Chatrchyan:2012cg}, and 
that of the $\ell_3$, where it is less than 2\%~\cite{Aad:2013izg}.
As a conservative estimate, which also allows a direct comparison between the
$3\ell$ channel and this channel, we set $\varepsilon = 0.21$ 
here also. This results in $\mathcal{Z} = 27\,\sigma$ in this
channel for the BP2, as seen in table~\ref{tab:StoB}, which is much
higher than the estimated $\mathcal{Z}$ for the same point in the 
$3\ell$ channel.

There is, however, a caveat here. As noted from
table~\ref{tab:BP2-result}, $B$ for the BP2 is much smaller than
$S$, resulting in a large $S/B$. In such a case, the approximate
expression for $\mathcal{Z}$, which assumes $S\ll B$,
is in principle not valid~\cite{Cowan:2010js}. While, for a
consistent treatment of the systematic uncertainties between the two
search channels, we retain this approximate expression for the $\muco$
channel also, we emphasize that the given $S/B$ values be considered
a much more accurate estimate of the strength of each channel at
the 14\tev\ LHC.

\begin{table}[tbp]
\centering
\begin{tabular}{|c|c|c|c|c|c|c|c|}
\hline         & ~~BP2 ~~        & ~~$W\gamma^*$  ~~    & ~~$Z\gamma^*$~~   & ~~$Wb\bar{b}$  ~~ \\
\hline  Cross section  (fb)                    &3.93 &246.9 & 10.0 & 3770.0\\
\hline  Cut efficiency                         & 0.050 & $5.3\times 10^{-5}$ & $3\times 10^{-5}$ & $1\times 10^{-6}$ \\
\hline  Effective cross section (fb)  & 0.197 & 0.013 & 0.0003 & 0.003 \\
\hline No. of events & 59.1 & 3.9 & 0.09 & 0.9\\
\hline
\end{tabular}
\caption{The backgrounds and the signal for the BP2 in the \muco\ search channel at the 14\,TeV LHC for $\mathcal{L}=300$\,fb$^{-1}$.}
\label{tab:BP2-result}
\end{table}

\begin{table}[tbp]
\centering
\begin{tabular}{|c||c|c||c|c|}
\hline  Point  & \multicolumn{2}{|c||}{$S/B$ in analysis}  &
\multicolumn{2}{|c|}{$\mathcal{Z}~(\sigma)$ in analysis} \\
\hline
\hline
  & $3\ell$  (SRZc region) &  $\muco$ & $3\ell$  (SRZc region) &
  $\muco$ \\
 \hline BP1 & 0.591 & 0.42 & 2.7 & 1.2 \\
 \hline BP2  &  0.436 & 15  & 2.0  & 27    \\
\hline
\end{tabular}
\caption{Results in the two analyses methods for the benchmark points.}
\label{tab:StoB}
\end{table}

\section{\label{sec:concl} Conclusions}

In this article we have discussed an $\mathcal{O}(1)$\gev\
neutralino DM in the NMSSM and its detectability 
at the LHC. Despite being very
light, such a singlino-like DM can generate thermal relic abundance of
the universe consistent with the PLANCK measurement, owing to 
the existence also of a singlet-like pseudoscalar, $A_1$, with a mass 
around twice the DM mass. A very light DM giving
the correct relic abundance is impossible to obtain in the MSSM, and
thus its detection will provide a clear indication of physics beyond
minimal SUSY. We have noted that the current direct and indirect
detection facilities have very poor detection prospects for such a
light DM with and hence its dedicated searches at the LHC can prove 
very crucial. We have therefore studied in depth the prospects for
its observation at the 14\,TeV LHC. 

By performing a through scan of the NMSSM parameter space, chosen
taking into account the analytical structure of the neutralino mass matrix, 
we found a variety of its configurations where an $\mathcal{O}(1)$\gev\ DM
can be obtained. We then carried out detector-level analyses of
two of the main production modes of such a DM. In both these
modes, the DM is produced in the decays of a higgsino-like heavier
neutralino, $\neut{2/3}$, which itself is produced in a pair with the lightest
chargino. The $\neut{2/3}$ then decays into either $Z + \mathrm{DM}$
or $A_1 + \mathrm{DM}$. The former 
channel, combined with the chargino decay, results in a ${\rm
trilepton} + \met$ final state, for which dedicated searches
are already being performed by the CMS and ATLAS collaborations. 
In the latter channel the final state comprises of a pair of
muons and a third lepton along with the DM. 

In the $\neut{2,3} \to \mathrm{DM} + A_1$ channel, the two muons that 
the very light $A_1$ decays into are highly collinear. 
Therefore, this channel can not be probed using the usual muon identification
criteria. For this reason, we
have adopted the technique of grouping the two muons together into a single
jet-like object by applying certain unconventional rigorous cuts. By
 implementing this method on two benchmark points from our 
scan, we have found that 
this channel can have a signal strength comparable to that of the
trilepton channel at the 14\,TeV LHC with $\mathcal{L} = 300\,{\rm
  fb}^{-1}$. In fact, for one of the two points, wherein the
BR($\neut{2,3} \to \mathrm{DM} + A_1$) is significant and the
 BR($A_1 \to \mu^+\mu^-$) is maximal, the obtained $S/B$ for this cannel is 
much larger than that for the trilepton channel. We thus emphasize that
dedicated searches in this channel may prove very crucial for the
discovery of a very light SUSY DM within a few years of the current 
LHC run.

\section*{Acknowledgments}

C.~Han is thankful to Peiwen Wu for useful discussions. This work is
supported by the Korea Ministry of Science, ICT and Future Planning, 
Gyeongsangbuk-Do and Pohang City for Independent Junior Research 
Groups at the Asia Pacific Center for Theoretical Physics. MP is also
supported by World Premier International Research Center Initiative 
(WPI Initiative), MEXT, Japan.

\bibliographystyle{utphysmcite}
\bibliography{Singlino}

\end{document}